\documentclass[11pt]{article}
\usepackage{moriond,epsfig}

\def\be{\begin{equation}}
\def\ee{\end{equation}}
\def\bea{\begin{eqnarray}}
\def\eea{\end{eqnarray}}

\newcommand{\met}{\mbox{$E_{T}\!\!\!\!\!\!\!/\,\,\,\,$}}

\begin{document}
%\vspace*{4cm}
\title{NON-SUSY SEARCHES AT THE TEVATRON}

\author{John Strologas }

\address{Department of Physics and Astronomy, University of New Mexico, \\Albuquerque,
New Mexico 87131, USA}

\maketitle\abstracts{
We present recent results from searches for new physics beyond supersymmetry
performed at the Tevatron accelerator at Fermilab.  The CDF and D$\O$ analyses
presented here utilized data of integrated luminosity up to 6 fb$^{-1}$.
We cover leptonic and bosonic resonances interpreted in the Randall-Sundrum graviton and 
new-boson models, rare final states, and the search for vector-like quarks.}

\section{Introduction}
The search for new phenomena beyond the weak-scale supersymmetry is a vital part of the Fermilab program.  Both CDF and D$\O$ experiments at the Tevatron collider actively look for signals not expected by the standard model (SM) or minimal supersymmetric models.  The searches can be sorted in three categories: a) searches for generic resonances that can be interpreted in several new-physics models b) searches for exotic combinations of final-state objects or abnormal kinematics  (not necessarily predicted by current theories) and c) model-dependent searches that test a particular theory.  We present here latest results from all these categories:  searches for new dilepton and diboson resonances (interpreted as gravitons and new gauge bosons), searches for anomalous $\gamma+\met$+X production, and searches for vector-like quarks.

\section{Search for New Resonances}

\subsection{Search for RS-gravitons}
\begin{figure}[!]
\begin{minipage}[!]{7.5cm}  
\begin{center}
\includegraphics[scale=.4]{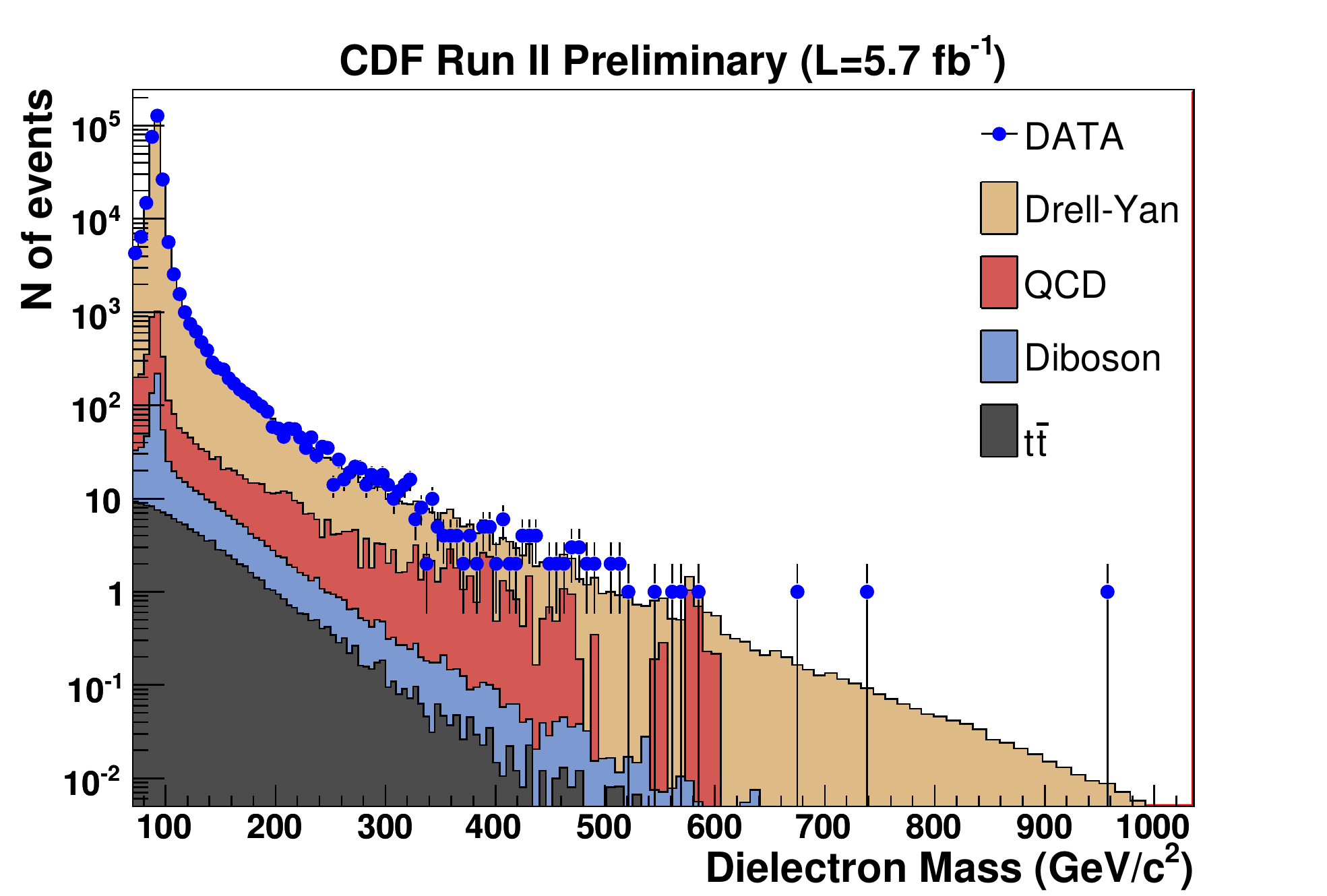}
\caption{Dielectron mass spectrum from the CDF search for RS-gravitons. \label{RSspectrum}}
\end{center}
\end{minipage}  
\hfill
\begin{minipage}[!]{7.5cm} 
\begin{center}
\vspace{0.3cm}
\includegraphics[scale=.4]{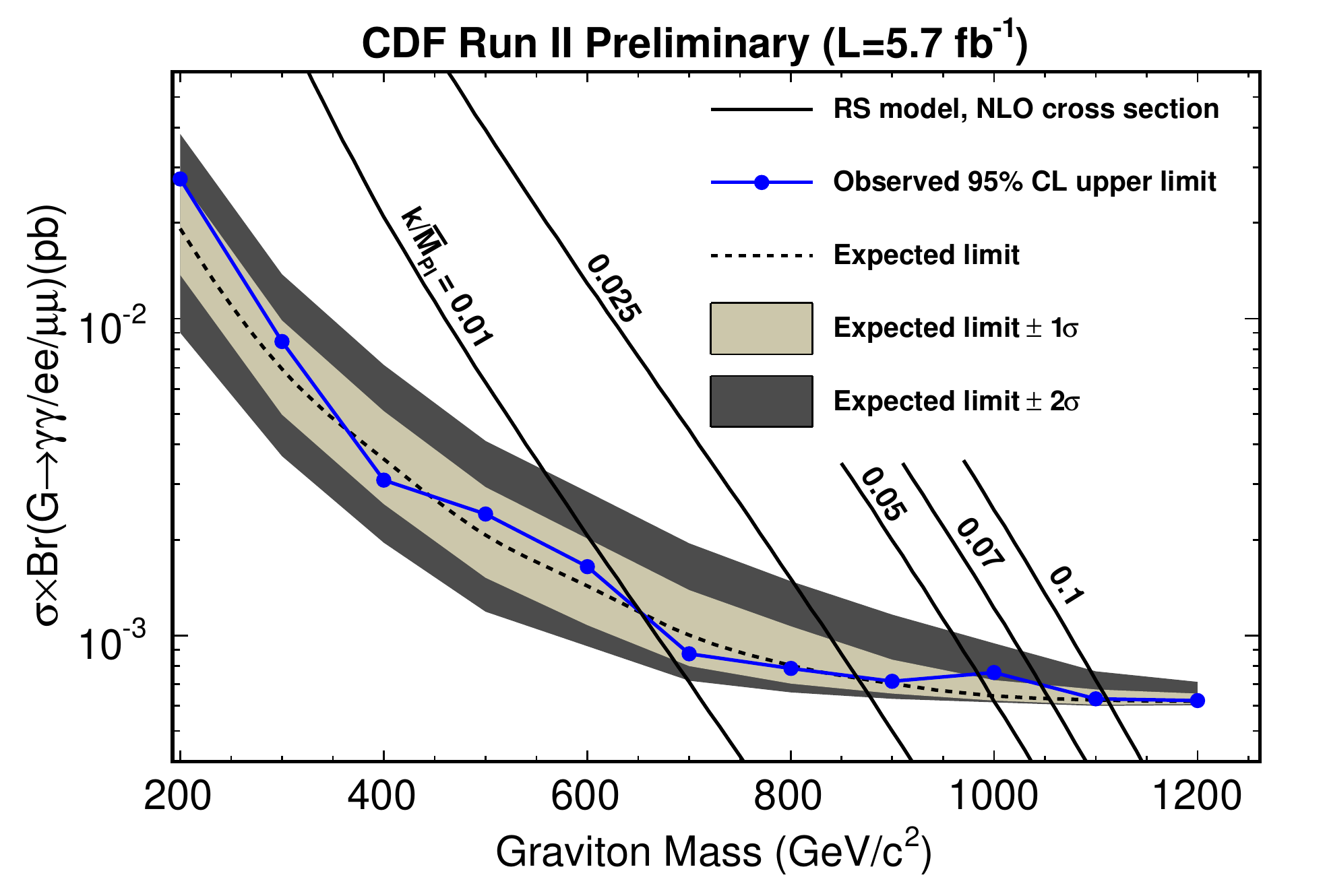}
\vspace{-0.7cm}
\caption{Exclusion plots from the CDF search for RS-gravitons decaying to $\gamma\gamma$ or $ee$ or $\mu\mu$. \label{RSlimit}}
\end{center}
\end{minipage}  
\hfill
\end{figure}

The Randall-Sundrum (RS) model offers an explanation for the hierarchy problem with the introduction of an extra dimension accessible only to gravity-carrying gravitons.  CDF completed a search for the RS-gravitons in the dielectron \cite{cdf_rs_ee} and dimuon \cite{cdf_rs_mm} channels using 5.7 fb$^{-1}$ of integrated luminosity.  Two electrons (muons) are selected, at least one with $E_T$ ($p_T$) above 20 GeV (GeV$/c$).  The leptons are isolated, i.e., the excess energy in a cone $\Delta R = \sqrt{\Delta\phi)^2+(\Delta\eta)^2}=0.4$ around each lepton is less than 10\% of the energy (momentum) of the electron (muon).  Cosmic-ray veto and $\gamma\rightarrow ee$ conversion removal are applied; no opposite charge is enforced.  Main backgrounds come from the Drell-Yan (DY) process, and QCD background associated with one real lepton and one "fake" lepton (i.e., jet (track) faking an electron (muon)).  Minor backgrounds come from diboson and $t\bar{t}$ processes.  The QCD background is determined with CDF data, with the application of a probability that a jet (track) fakes an electron (muon) on events with one identified lepton.  All other backgrounds are estimated with Monte Carlo simulations (MC) absolutely normalized to the next-to-leading order cross sections, the data luminosity, lepton-identification scale factors and trigger efficiencies.  The dielectron and dimuon mass spectra are consistent with expectation.  At the same time, the highest-dielectron-mass event ever observed is detected ($M_{ee}=960$ GeV/$c^2$).  The probability that at least one event is observed with a dielectron mass at least that high is 4\%.  The results are interpreted in the RS-graviton model, combined with a previous diphoton \cite{cdf_rs_gg} graviton search.  For the RS-model parameter $k/\overline{M}_{Pl}=0.1$, RS-gravitons with mass less than 1111 GeV$/c^2$ are excluded at 95\% confidence level (CL). Figure \ref{RSspectrum} shows the dielectron spectrum and Figure \ref{RSlimit} shows the cross-section uper limit as a function of the graviton mass, along with theoretical cross sections for several values of $k/\overline{M}_{Pl}$.

\subsection{Search for $Z'$}
\begin{figure}[!]
\begin{minipage}[!]{7.5cm}  
\begin{center}
\includegraphics[scale=.4]{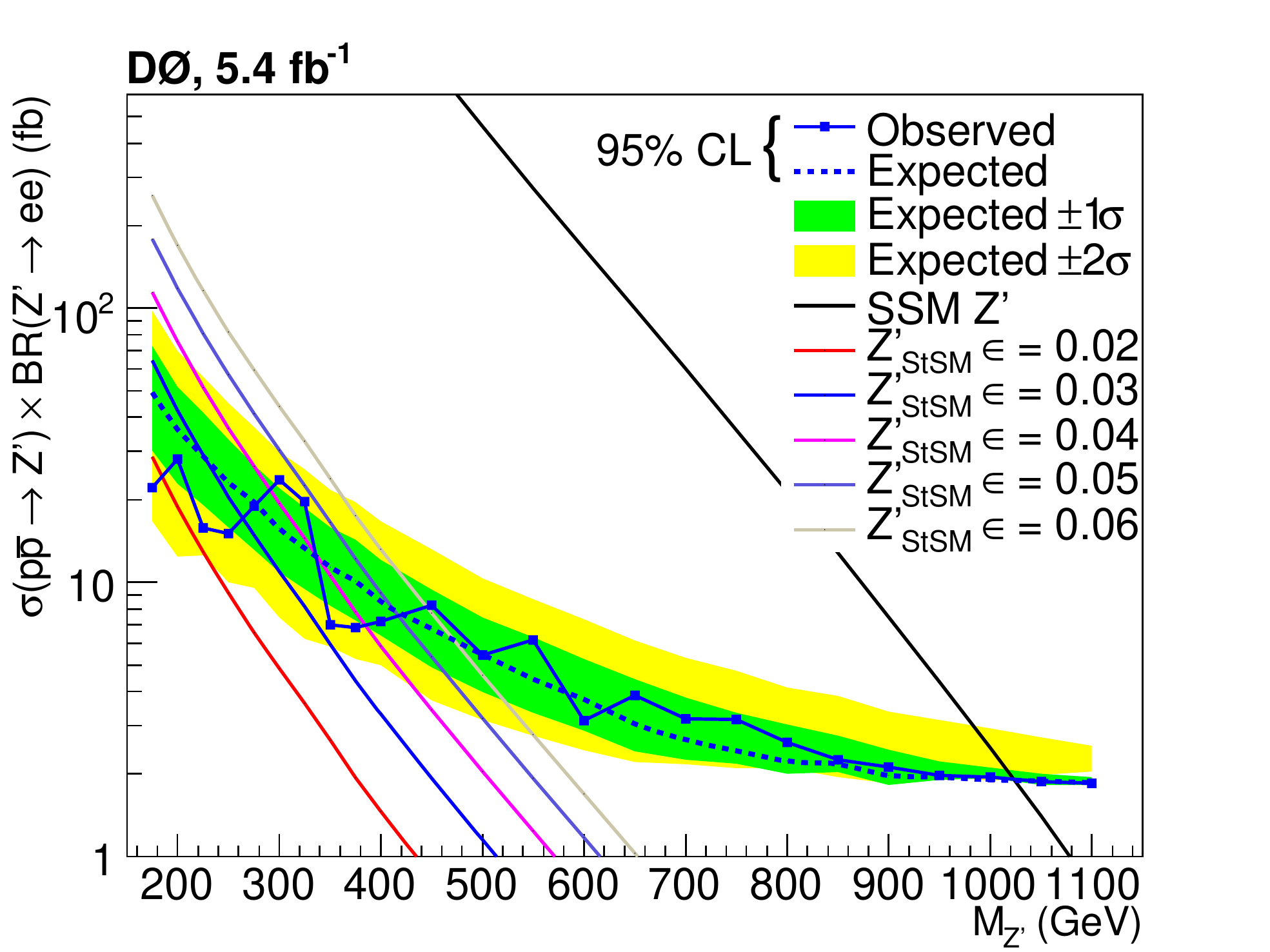}
\caption{Exclusion plot from the D$\O$ search for $Z'$ decaying to two electrons. \label{Zprime_ee}}
\end{center}
\end{minipage}  
\hfill
\begin{minipage}[!]{7.5cm} 
\begin{center}
\vspace{0.3cm}
\includegraphics[scale=.4]{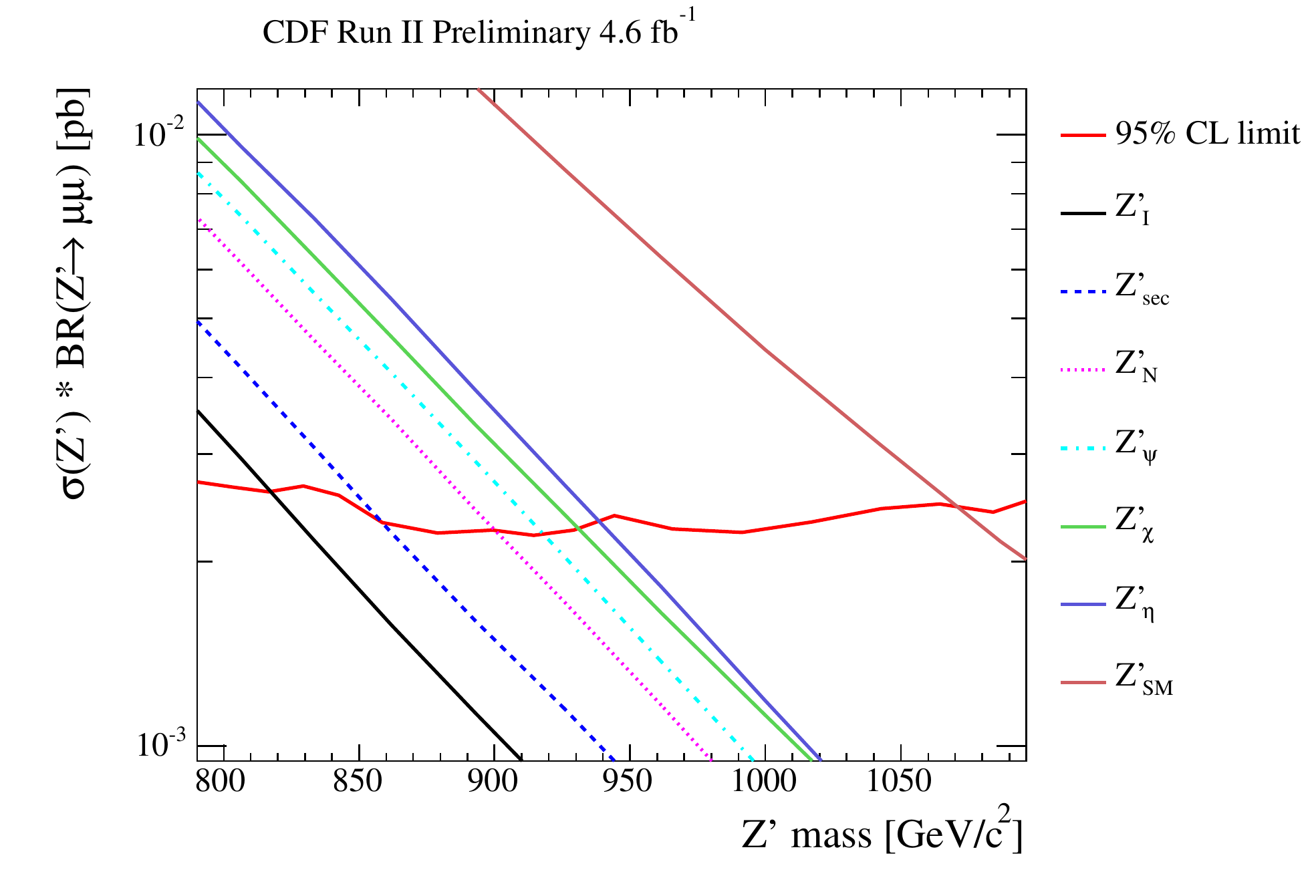}
\vspace{-0.22cm}
\caption{Exclusion plot from the CDF search for $Z'$ decaying to two muons. \label{Zprime_mm}}
\end{center}
\end{minipage}  
\hfill
\end{figure}

New gauge-bosons are predicted by Grand Unifying Theories and would appear as resonances at hadron colliders.  D$\O$ recently searched for a new gauge boson $Z'\rightarrow ee$ using 5.4 fb$^{-1}$ of integrated luminosity \cite{d0_zp_ee}.  Two electrons with $p_T>25$ GeV$/c$ are selected and no opposite charge is required.  The electrons are isolated: the excess energy in a cone $\Delta R = 0.4$ around the electron is less than 7\% of its energy; additional track-based isolation is applied.  The backgrounds are identical to the CDF dielectron search.  The QCD background is estimated using a dielectron mass shape from D$\O$ data that fail the electron identification selection and fitting in the $Z$ resonance along with the DY background.  The remaining minor SM backgrounds (diboson, $t\bar{t}$) are estimated using MC absolutely normalized.  The dielectron spectrum is consistent with the one expected by the SM.  For a $Z'$ interacting with SM couplings, $Z'$ masses below 1023 GeV/$c^2$ are excluded at 95\% CL.  Figure \ref{Zprime_ee} shows the exclusion limits for several $Z'$ models.

CDF recently published a $Z'\rightarrow \mu\mu$ search using 4.6 fb$^{-1}$ of integrated luminosity \cite{cdf_zp_mm}.  The two muons are above 30 GeV$/c$, a cosmic-ray veto is applied and no opposite charge is required.  The isolation requirements and the electroweak/QCD background estimations are the same as in the CDF RS-graviton search using dimuons.  The mass spectrum is consistent with the SM; the SM-like $Z'$ allowed masses are above 1071 GeV/$c^2$ at 95\% CL.  Figure \ref{Zprime_mm} shows the exclusion limits for several $Z'$ models.

\subsection{Search for $W'$}
\begin{figure}[!]
\begin{minipage}[!]{7.5cm}  
\begin{center}
\includegraphics[scale=.35]{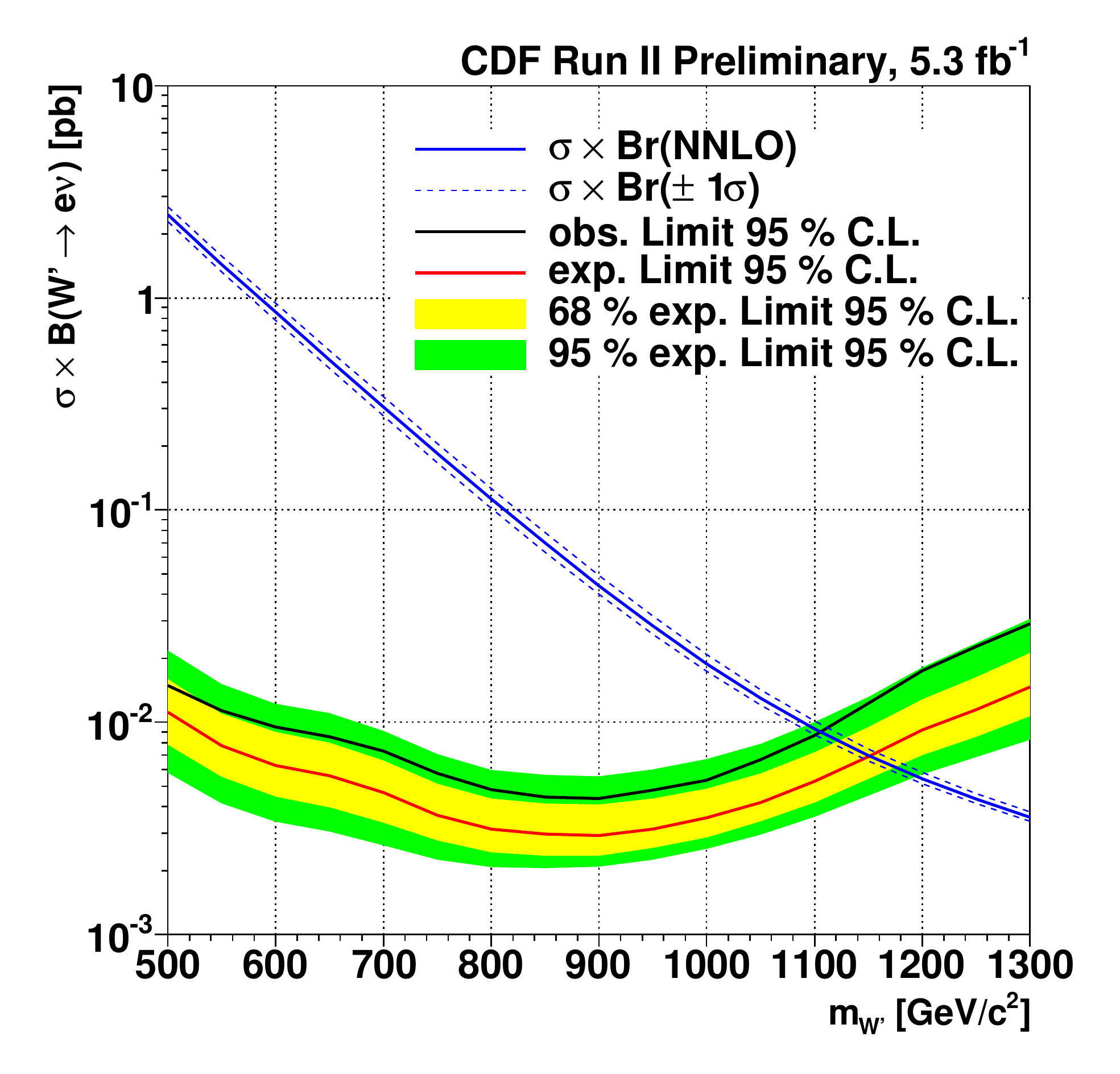}
\caption{Exclusion plot from CDF search for $W'$ decaying to an electron and a neutrino. \label{Wprime_en}}
\end{center}
\end{minipage}  
\hfill
\begin{minipage}[!]{7.5cm} 
\begin{center}
\vspace{0.3cm}
\includegraphics[scale=.4]{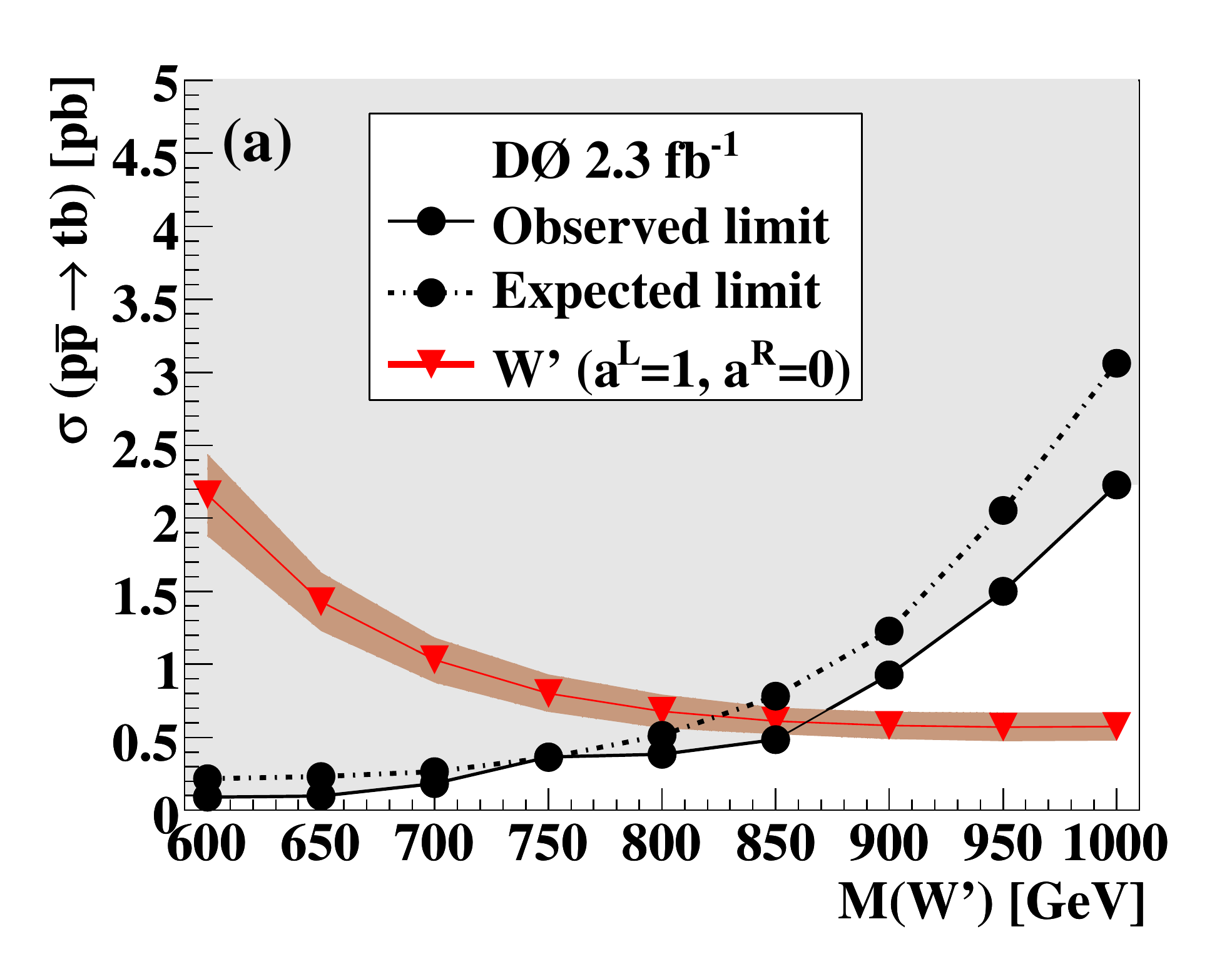}
\vspace{-0.27cm}
\caption{Exclusion plot from the D$\O$ search for $W'$ decaying to a top and bottom quarks. \label{Wprime_tb}}
\end{center}
\end{minipage}  
\hfill
\end{figure}

The leptonic decay of the $W'$ boson will lead to a lepton and missing transverse energy ($\met$) coming from the undetected neutrino.  CDF completed a 5.3 fb$^{-1}$ search for $W'$ decaying to electron and a neutrino \cite{cdf_wp_ne}.  One isolated electron with $E_T>25$ GeV and $\met>25$ GeV are required.  Given that the two objects have to have similar energy, $0.4<E_T/\met<2.5$ is required.  Events with two electrons are rejected.  The main SM background is coming from the SM $W$ decaying to the identical final state, $W/Z$ decaying to taus with at least one leptonic decay and QCD multijets with fake leptons and fake or real $\met$.  The electron-neutrino transverse-mass spectrum is in good agreement with expectation.  Assuming $W'$ with SM left-right symmetric couplings, $W'$ with a mass less than 1.12 TeV/$c^2$ are excluded at 95\% CL.  Figure \ref{Wprime_en} shows this exclusion limit.

In some models the new gauge boson couples stronger to the third generation.  D$\O$ performed a 2.3 fb$^{-1}$ analysis looking for $W'$ decaying to a top and a bottom quark with a final state of two bottom quarks, a lepton, and a neutrino \cite{d0_wp_tb}.  The events selected include one $e(\mu)$ with $p_T$ above 15/20 GeV$/c$.  Events are segregated based on jet multiplicity (2, 3, or 4 jets) counting jets with $E_T>15$ GeV (leading jet above 25 GeV) and based on $b$-tagging (1 or 2 tags).  The total invariant mass is required to be above 400 GeV/$c^2$.  Main SM backgrounds include $W$+jets and $t\bar{t}$ production.  Minor backgrounds come from QCD multijet and $Z$+jets.  Lower mass limits are presented for different combinations of left-handed and right-handed couplings and range from 863 to 916 GeV/c$^2$ at 95\% CL.  Figure \ref{Wprime_tb} shows the exclusion limits for a $W'$ with SM couplings.

\subsection{Diboson Resonances}
D$\O$ searched for $WW$ and $WZ$ resonances using 4.1 fb$^{-1}$ (trileptons) or 5.4 fb$^{-1}$ (lepton(s)+jets) of data \cite{d0_diboson}.  Three selections were used: either a $W$ decays leptonically or the $Z$ decays leptonically or both bosons decay leptonically, leading to 1-lepton+jets, 2-leptons+jets, or 3-lepton final signatures respectively.  The leptons and jets are required to be above 20 GeV, where the $\met$ for the three channels is required to be $\met>20$, $\met<50$, and $\met>30$ GeV respectively.  In this analysis boosted $W/Z$ decays to two jets could be reconstructed as one jet.  Main backgrounds are $W/Z$+jets and diboson. The observations are consistent with the SM expectation.  Due to higher backgrounds and under the assumption of lower branching fractions to leptons, the limits interpreted in the RS-graviton model or in $W'$ are weaker than the respective direct dilepton ones ($M(W')>690$ GeV/$c^2$and $M(G)>754$ GeV/$c^2$ at 95\% CL).
\section{Search for Exotic Signatures}

\subsection{Search for exotic production of $\gamma$+jet}
CDF performed a 4.8 fb$^{-1}$ analysis \cite{cdf_exotic_1} searching for abnormal generation of photon+jet, requiring the photon $E_T$ to be above 30 GeV and the jet above 15 GeV.  The photon and jet were separated so that $\Delta\phi({\rm photon,jet})>0.4$ rad.  Four channels were investigated ($>=1$ and $>=2$ jets, with and without requiring $\met>20$ GeV).  Main backgrounds are the SM production of $\gamma$, $\gamma\gamma$, charged-lepton, QCD multijets, cosmic rays and beam halo. Good agreement is observed in several kinematic distributions and no hint of new physics is seen.

\subsection{Search for exotic production of lepton+$\gamma$+$\met$+$b$-quark}
CDF also performed a 6 fb$^{-1}$ search \cite{cdf_exotic_2} for abnormal production of lepton+$\gamma$+$\met$+$b$-quark requiring lepton $E_T>20$ GeV, $\met>20$ GeV, photon $E_T>12$ GeV, and a $b$-tagged jet with $E_T>20$ GeV.  Main backgrounds are $t\bar{t}+\gamma$ and $W+\gamma$+jets.  No new physics is seen, and this analysis yields an impressive measurement of SM $t\bar{t}+\gamma$ cross section of $\sigma=0.18 \pm 0.07 {\rm (stat.)} \pm 0.04 {\rm (sys.)} \pm 0.01 {\rm (lum.)}$ pb.

\section{Search for Vector-like Quarks}
D$\O$ performed a 5.4 fb$^{-1}$ search for vector-like quarks that would manifest themselves in $W/Z$ + 2 jets \cite{d0_vl_quarks}.  For the $Z$+jets channel two electrons or muons above 20 GeV with $p_T(\rm{dilepton})>100$ GeV/$c$, at least two jets above 20 GeV (leading $>100$ GeV), $\met<50$ GeV and $70<M_{\ell\ell}<110$ GeV/$c^2$ are required.  Main background comes from SM $Z$+jets.  The mass of the vector-like quark must be above 430-551 GeV/$c^2$ (depending on the couplings) at 95\% CL.  For the $W$+jets channel one electron or muon is selected with $p_T>50$ GeV/$c$, same jet requirement, $\met>50/40$ GeV and $(2M_T^Wc^2+\met)>80$ GeV.  The main backgrounds come from SM $W$+jets, top quarks, QCD multijets, $Z$+jets, and diboson.  The mass of the vector-like quark must be above 403-693 GeV/$c^2$ (depending on the couplings) at 95\% CL. 

\section*{References}
\begin{thebibliography}{99}

\bibitem{cdf_rs_ee} T. Aaltonen {\em et al}.\ (CDF Collaboration), arXiv:1103.4650 [hep-ex] (2011).
\bibitem{cdf_rs_mm} T. Aaltonen {\em et al}.\ (CDF Collaboration), CDF public note 10479 (2011).
\bibitem{cdf_rs_gg} T. Aaltonen {\em et al}.\ (CDF Collaboration), Phys.\ Rev.\ D {\bf 83}, 011102(R) (2011).
\bibitem{d0_zp_ee}  V. Abazov {\em et al}.\ (D0 Collaboration), Phys.\ Lett.\ B {\bf 695}, 88 (2011).
\bibitem{cdf_zp_mm} T. Aaltonen {\em et al}.\ (CDF Collaboration), Phys.\ Rev.\ Lett.\ {\bf 106}, 121801 (2011).
\bibitem{cdf_wp_ne} T. Aaltonen {\em et al}.\ (CDF Collaboration), Phys.\ Rev.\ D {\bf 83}, 031102 (2011).
\bibitem{d0_wp_tb} V. Abazov {\em et al}.\ (D0 Collaboration), Phys.\ Lett.\ B {\bf 699}, 145 (2011).
\bibitem{d0_diboson} V. Abazov {\em et al}.\ (D0 Collaboration), Phys.\ Rev.\ Lett.\ {\bf 107}, 011801 (2011).
\bibitem{cdf_exotic_1} T. Aaltonen {\em et al}.\ (CDF Collaboration), CDF public note 10355 (2011).
\bibitem{cdf_exotic_2} T. Aaltonen {\em et al}.\ (CDF Collaboration), CDF public note 10270 (2011).
\bibitem{d0_vl_quarks} V. Abazov {\em et al}.\ (D0 Collaboration), Phys.\ Rev.\ Lett.\ {\bf 106}, 081801 (2011).

\end{thebibliography}
\end{document}